\documentclass{svproc}
\usepackage{url}

\usepackage{graphicx}

\begin{document}
\mainmatter              
\title{CP violating triple product asymmetries in charm decays}
\titlerunning{CP violating triple product asymmetries in charm decays}
%
\author{S.~Bahinipati\\
(On behalf of the Belle Collaboration)}
\authorrunning{S. Bahinipati (On behalf of the Belle Collaboration)} 
%
%
\institute{Indian Institute of Technology Bhubaneswar, Bhubaneswar, India\\
Email:\email{seema.bahinipati@iitbbs.ac.in}
}

\maketitle              

\begin{abstract}

$CP$ violation asymmetry is expected to be very small within the Standard Model. Observation of $CP$ violation in charm sector would hint towards New Physics. Time-reversal asymmetry is sensitive to $CPV$ via the $CPT$ Theorem, and is a clean and alternative way to search for CP violation in charm decays. This is achieved by the use of triple-product correlations in 4-body charm meson decays. The results from previous searches at BaBar and LHCb Collaborations and the latest results from Belle are reported. The most recent result from Belle performed in $D^{0} \to K^{+}K^{-}\pi^{+}\pi^{-}$ decays using kinematic asymmetry  variables is the first measurement of its kind.

\end{abstract}

\section{Introduction}
\label{intro}

$CP$ asymmetry in charm decays is expected to be quite small in the Standard Model (SM), of the order of $10^{-3}$~\cite{Ref1}. Hence, any deviation from this predicted small asymmetry in charm decays would hint towards physics beyond SM~\cite{Ref2}. Using the large samples of charm decays collected at B-factories such as BaBar, Belle and LHCb collaborations, the search for $CP$ violation in charm decays. Time-reversal (T) asymmetry is sensitive to CP violation via the CPT theorem~\cite{Ref3}. Hence, these collaborations performed T-asymmetry searches using multi-body charm decays.

\section{Previous Results from BaBar}
The BaBar collaboration used triple-product correlations to search for $CP$ violation in $D^{+} \to K^{+} K_{S}^{0}\pi^{+}\pi^{-}$~\cite{Refbb1} and $D_{s}^{+} \to K^{+} K_{S}^{0}\pi^{+}\pi^{-}$~\cite{Refbb1} decays using $520~fb^{-1}$ of data recorded by the BABAR detector at the PEP-II asymmetric-energy collider operating at center of mass energies near 10.6 GeV. A kinematic triple product that is odd under time reversal using the vector momenta of the final state particles in the $D_{(s)}^{+}$ rest frame is defined as: $C_{T} = {\vec{p_{K^{+}}}} \cdot  {\vec{p_{\pi^{+}}}} \times {\vec{p_{\pi^{-}}}}$.

\begin{equation}
A_{T} = \frac { \Gamma(C_{T}>0) - \Gamma(C_{T}<0)} { \Gamma(C_{T}>0) + \Gamma (C_{T}<0)}, 
\end{equation}
\begin{equation}
{\bar{A}}_{T} = \frac { \Gamma(-{\bar{C_{T}}}>0) - \Gamma(-{\bar{C_{T}}}<0)} { \Gamma(-{\bar{C_{T}}}>0) + \Gamma (-{\bar{C_{T}}}<0)}, 
\end{equation}

where $\Gamma$ is the decay rate.
The CP violating observable is given by the difference of these two asymmetries
\begin{equation}
a_{T-odd}^{CP} = \frac{1}{2} (A_{T}  - {\bar{A}}_{T})
\end{equation}
\begin{equation}
a_{T-odd}^{CP} = (-12.0 \pm 10.0 \pm 4.6) \times 10^{-3} 
\end{equation}

\begin{equation}
a_{T-odd}^{CP} = (-13.6 \pm 7.7 \pm 3.4) \times 10^{-3} 
\end{equation}

Studies suggest that these correlations can be used to probe C and P symmetries as well~\cite{Refbevan}. The BaBar Collaboration reinterpreted their results on CP violation using triple-product correlations in light of these studies~\cite{Refckm14}.

\section{Previous Results from LHCb}
\label{lhcb}
\begin{figure}[h]
\includegraphics[scale=0.5]{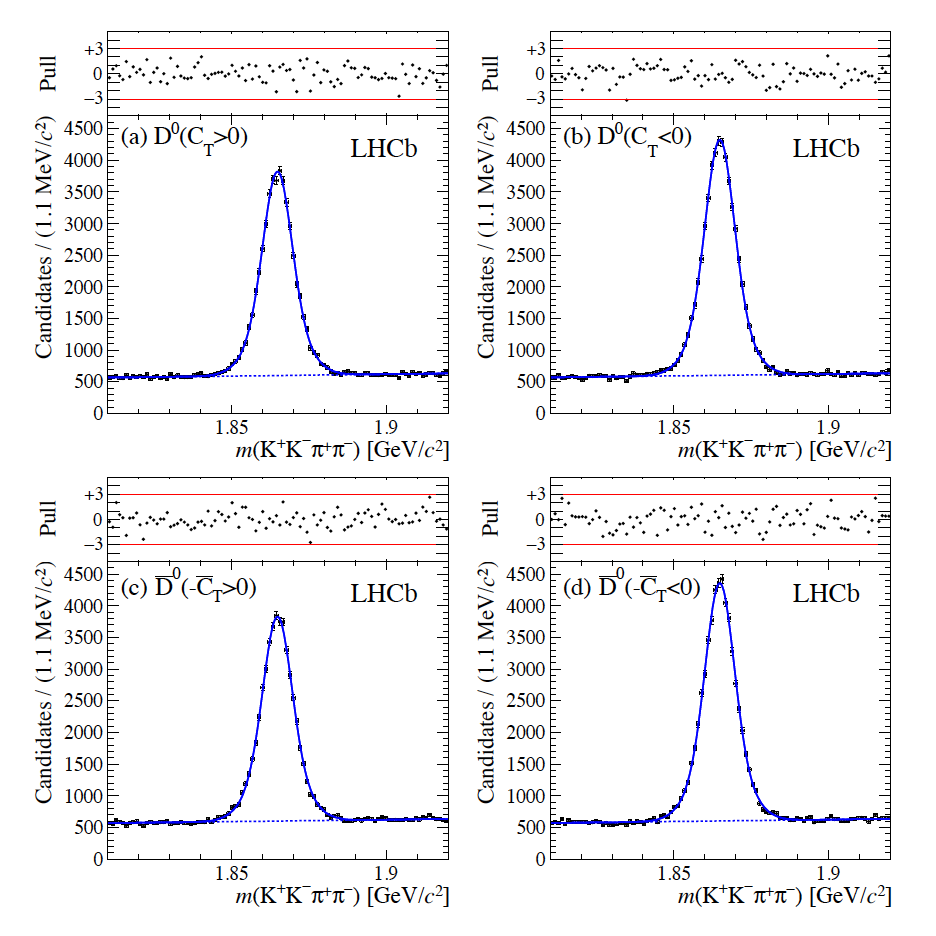}
\caption{$D^{0}$-candidate mass distributions depending on flavour and the value of $C_{T}$.
In each case, the result of the fit is overlaid as a solid curve, while a dashed line represents
the background. The distribution of normalised residual (pull), defined as the difference
between the fit result and the data, divided by the data uncertainty, is shown above each
data distribution.}
\label{fig:lhcb1}       
\end{figure}

The LHCb collaboration used triple-product correlations to search for $CP$ violation in $D^{0} \to K^{+} K^{-}\pi^{+}\pi^{-}$ decays~\cite{Reflhcb}, using $3~fb^{-1}$ of data recorded by the LHCb detector in 2011 and 2012, when selecting them through partial reconstruction of semileptonic
decays of the B meson ($B \to D^{0} \mu^{-}X$, where $X$ indicates any system composed of
charged and neutral particles). he charge of the muon is used to identify the flavour
of the $D^{0}$ meson. The triple-product is defined by using the momenta
of three out of the four daughters in the $D^{0}$ rest frame, $C_{T} = {\vec{p_{K^{+}}}} \cdot  {\vec{p_{\pi^{+}}}} \times   {\vec{p_{\pi^{-}}}} $. The distributions of the  $D^{0}$meson candidates in four different
regions defined by $D^{0}$ flavour and $C_{T}$ value being greater or less then zero are shown in Fig.~\ref{fig:lhcb1}.

\begin{equation}
A_{T} = \frac { \Gamma(C_{T}>0) - \Gamma(C_{T}<0)} { \Gamma(C_{T}>0) + \Gamma (C_{T}<0)}, 
\end{equation}
\begin{equation}
{\bar{A}}_{T} = \frac { \Gamma(-{\bar{C_{T}}}>0) - \Gamma(-{\bar{C_{T}}}<0)} { \Gamma(-{\bar{C_{T}}}>0) + \Gamma (-{\bar{C_{T}}}<0)}, 
\end{equation}

where $\Gamma$ is the decay rate.
The CP violating observable is given by the difference of these two asymmetries
\begin{equation}
a_{T-odd}^{CP} = \frac{1}{2} (A_{T}  - {\bar{A}}_{T})
\end{equation}

The LHCb Collaboration performed three different searches for CP violation: a time-integrated measurement, a
measurement as a function of phase-space region and a measurement as a function of $D^{0}$-decay-time region. The result of the integrated measurement is $a_{CP}^{T-odd} = (1.8 \pm 2.9 \pm 0.4) \times 10^{-3}$, where the two uncertainties are statistical and systematic, respectively. No significant deviation from zero is observed here and in the measurement as a function of phase-space region and a measurement as a function of $D^{0}$-decay-time region as well.

\section{Latest Results from Belle}
\label{belle}

\subsection{$D^{0} \to K_{s} \pi^{+}\pi^{-}\pi^{0}$}
\label{bellesec:1}

Belle reported the first measurement of the T-odd moments in the decay in $D^{0} \to K_{s} \pi^{+}\pi^{-}\pi^{0}$ using $966~fb^{-1}$ of data recorded by the Belle experiment at the KEK-B asymmetric $e^{+}-e^{-}$ collider~\cite{Refbe1}. The triple products are defined as $C_{T} = {\vec{p_{K_{S}^{0}}}} \cdot  {\vec{p_{\pi^{+}}}} \times   {\vec{p_{\pi^{-}}}} $ for $D^{0}$ events and ${\bar{C}}_{T} = {\vec{p_{K_{S}^{0}}}} \cdot  {\vec{p_{\pi^{+}}}} \times   {\vec{p_{\pi^{-}}}} $ for ${\bar{D^{0}}}$ events. In order to determine $a_{T-odd}^{CP}$, the sample is divided into four categories using the $C_{T}$ values and $\pi_{slow}$ charge. The signal yields are calculated via a two-dimensional unbinned maximum likelihood fit to $D^{0}$ mass and $\Delta m$, which is the mass difference between $D^{*}$ and $D^{0}$ mass.

\begin{figure}[h]
\includegraphics[scale=0.5]{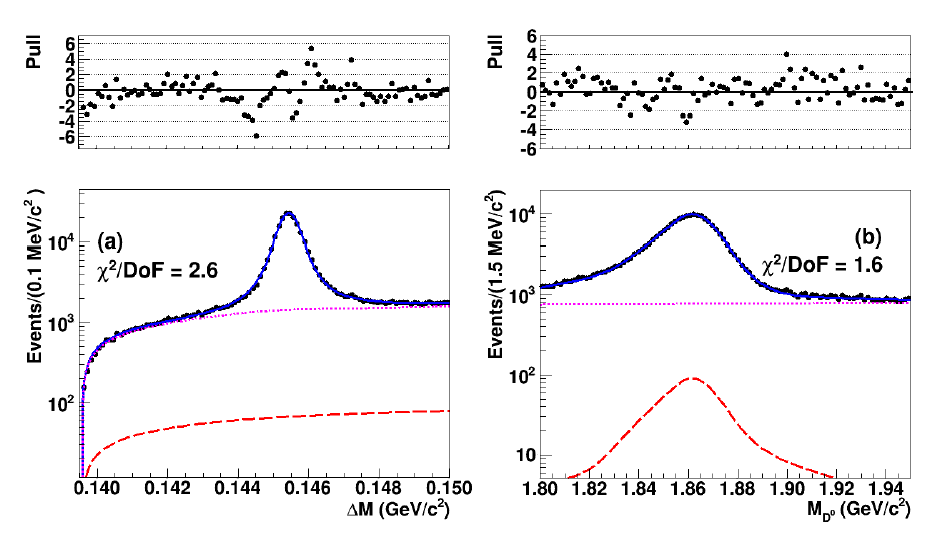}
\caption{The signal-enhanced logarithmic distributions of (a)$\Delta m$ (b) $M_{D^{0}}$ for $D^{0}$ with $C_{T} > 0$. The black points with error bars are the data points and the solid blue curve is the projection of the total signal and background components. The dotted magenta and dashed red curves indicate combinatorial and random+ slow backgrounds, respectively. The normalized residuals (pulls) and the $\chi^{2}/DoF$, where $DoF$ is the number of degrees of freedom is shown above each plot.}
\label{fig:b1}       
\end{figure}

The signal-enhanced $D^{0}$ mass and $\Delta m$ distributions of the data for the four categories  along with the fit projections are shown in Fig.~\ref{fig:b1}. The total signal yield obtained from is $744509\pm1622$ and the asymmetry $a_{T-odd}^{CP}$ is found to be $(-0.28 \pm 1.38) \times 10^{-3} $ , where the error is statisical error.
The search was aslo performed in $D^{0} \to K_{s} \pi^{+}\pi^{-}\pi^{0}$ phase space into nine exclusive bins according to the intermediate resonance contributions. The $A_{T}$ and $a_{T-odd}^{CP}$ are found to be consistent with zero.

The sources of systematic uncertainties are the signal and background models, efficiency dependence on $C_{T}$, $C_{T}$ resolution, and potential fit bias. The dominant contribution comes from modelling the signal and background PDFs.

\subsection{$D^{0} \to K^{+} K^{-}\pi^{+}\pi^{-}$}
\label{bellesec:2}
The most recent result from Belle is the search for CP Violation with kinematic asymmetries in the
$D^{0} \to K^{+} K^{-}\pi^{+}\pi^{-}$~\cite{Refbe2}. The advatange of using kinematic asymmetries is that they probe the rich variety of interfering contributions in the decay where CP-violation phases could be hiding. The kinematic asymmetries are defined as:

\begin{equation}
A_{X} = \frac { \Gamma(X>0) - \Gamma(X<0)} { \Gamma(X>0) + \Gamma (X<0)}, 
\end{equation}
where $\Gamma$ is the decay rate and $X$ is the kinematic function for the CP-conjugate decay.
\begin{equation}
a_{X}^{CP} = \frac{1}{2} A_{X} - \eta_{X}^{CP}{\bar{A}}_{\bar{X}}
\end{equation}
where $\eta_{X}^{CP}$ is the product of the CP eigenvalues of the interfering final states related with X. The kinematic asymmetries for $\eta_{X}^{CP} = +1 (-1)$ are the CP-violating T-odd (even) correlations, which are related with the real (imaginary) part of the interference contribution of the decay. Since these kinematic asymmetries are proportional to sin (weak phase), a non-zero value corresponds to CP-violation.

 The recent Belle result measures the kinematic asymmetry for six different kinematic functions, where five are measured for the first time. Five kinematic functions are constructed with angle $\theta_{1}$, $\theta_{2}$ and $\phi$, where $\theta_{1}$, $\theta_{2}$ are the helicity angles of the $K^{+}K^{-}$ and $\pi^{+}\pi^{-}$ systems against the positive charged particles respectively and $\phi$ is the angle between the decay planes of those systems. The $\eta_{X}^{CP}$ values for the six kinematic functions  $cos~\phi$, $sin~\phi$, $sin~2\phi$, $cos~\theta_{1} cos~\theta_{2} cos~\phi$,  $cos~\theta_{1} cos~\theta_{2} sin~\phi$ and  $C_{T}$ are $+1$,  $-1$,  $-1$,  $+1$,  $-1$ and  $-1$ respectively.

\begin{figure}[h]
\includegraphics[scale=0.45]{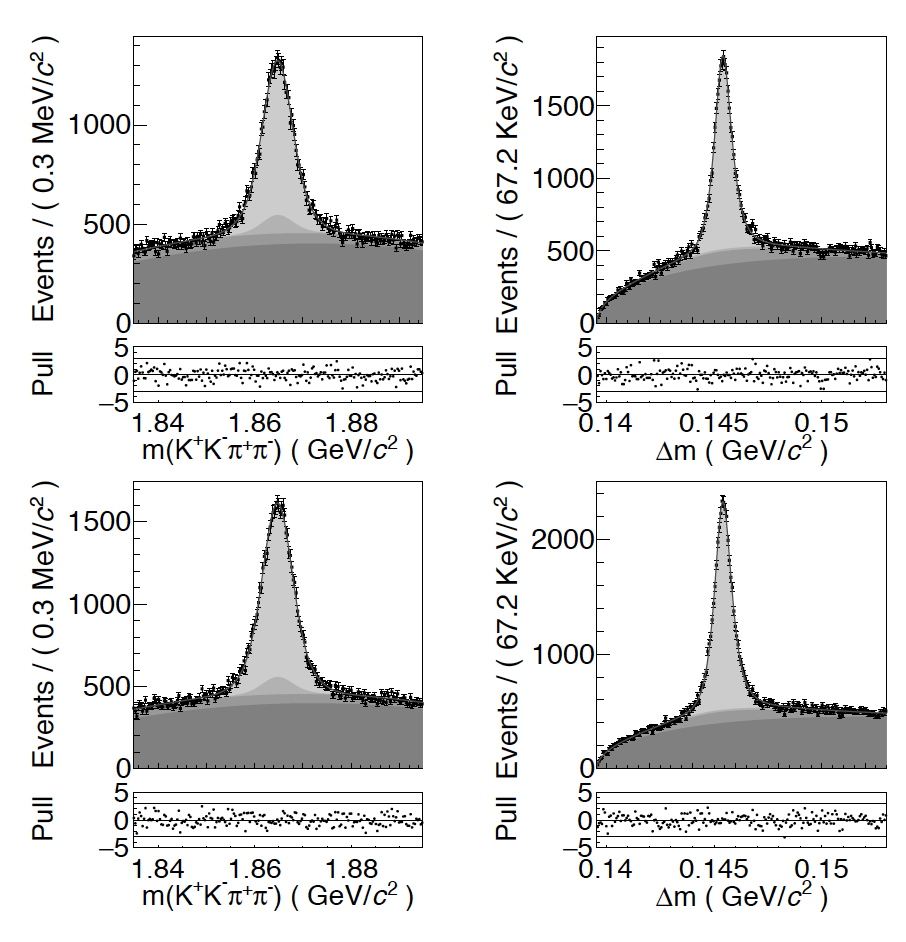}
\caption{Two-dimensional fit results, distributions and pull of the data sub-samples projected on the observables $K^{+} K^{-}\pi^{+}\pi^{-}$ mass and $\Delta m$. Top (bottom) histograms show the $D^{0}(sin~2\phi >0)$ and $D^{0}(sin~2\phi <0)$ sub-sample. The shaded regions are stacked upon each other and show, from lowest to highest, the combinatorial, partial-$D^{*}$, random$\pi_{s}$, and signal components. The lower plots show pulls for for the fit and the unlabeled horizontal lines indicate $\pm3$ and 0.}

\label{fig:b2}       
\end{figure}

The CP-violating kinematic asymmetry is calculated with the yield of the signal events for each flavor of $D^{0}$ and each sign of the relevant kinematic function. To do this, we perform four separate fits to the data for each kinematic function. The fits are two-dimensional extended maximum likelihood fit to the reconstructed $D^{0}$ mass and $\Delta m$. 
One model is included for all fits, and the model contains components describing signal, random slow pion, partial $D^{*}$ and combinatorial events. As an example of a set of fits used to determine the CP-violating asymmetry,  Fig.~\ref{fig:b2} shows a typical fit, showing the separate fit results for positive and negative $sin~2\phi$ for$ D^{0}$ samples.

 The asymmetries for the different kinematic variables defined for the decay, are found as folllows:
\begin{equation}
a_{cos~\phi}^{CP} = (3.4 \pm 3.6 \pm 1.7) \times 10^{-3} 
\end{equation}

\begin{equation}
a_{sin~\phi}^{CP} = (5.2 \pm 3.7 \pm 1.7) \times 10^{-3} 
\end{equation}

\begin{equation}
a_{sin~2\phi}^{CP} = (3.9 \pm 3.6 \pm 1.7) \times 10^{-3} 
\end{equation}

\begin{equation}
a_{cos~\theta_{1} cos~\theta_{2} cos\phi}^{CP} = (-0.2 \pm 3.6 \pm 1.6) \times 10^{-3}
\end{equation}

\begin{equation}
a_{cos~\theta_{1} cos~\theta_{2} sin\phi}^{CP} = (0.2 \pm 3.7 \pm 1.6) \times 10^{-3}
\end{equation}

\begin{equation}
a_{C_{T}}^{CP} = (5.0 \pm 3.7 \pm 1.6) \times 10^{-3} \\
\end{equation}
where the two uncertainties are statistical and systematic, respectively. The kinematic asymmetries are consistent with CP conservation and further constrain new physics models for the first time.

\section{Summary and Outlook}
\label{so}
In multi-body particle decays, the triple-product correlations provide an unique alternative and complementary handle to search for CP violation. LHCb has searched for CPV in $D^{0} \to K^{+} K^{-}\pi^{+}\pi^{-}$ decays. BaBar has searched for CPV in $D^{+} \to K^{+} K_{S}^{0}\pi^{+}\pi^{-}$ and $D_{s}^{+} \to K^{+} K_{S}^{0}\pi^{+}\pi^{-}$ decays. No sign of CPV was seen using T-odd correlations. The first measurement of T-odd moments was performed at Belle using $D^{0} \to K_{s} \pi^{+}\pi^{-}\pi^{0}$ decays. No CPV was observed. Furthermore, at Belle, CPV kinematic asymmetries was studied for the first time in $D^{0} \to K^{+} K^{-}\pi^{+}\pi^{-}$ decays. No CPV was observed; however, kinematic asymmetries further constrained NP models for the first time. Since the systematic uncertainties are small, triple product correlation is a promising topic to be pursued with large data samples, expected at Belle II and LHCb upgrade.



\end{document}